\documentclass[showpacs,amsmath,amssymb,prb,superscriptaddress,,reprint]{revtex4}
\usepackage{graphicx}
\usepackage{multirow}
\usepackage{textcomp}
\usepackage{color} 

\usepackage[colorlinks,plainpages=false,linkcolor=blue,urlcolor=blue,citecolor=blue,pdfpagemode=UseNone,pdfstartview=FitBH]{hyperref}

\begin{document}
\DeclareGraphicsExtensions{.eps, .jpg, .pdf}
\input epsf
\title{
Anisotropic compression in  the high pressure regime of pure and Cr-doped vanadium dioxide
}
\author {Matteo Mitrano\footnote{Current address:  Max Planck Research Dept. for Structural Dynamics, Notkestr. 85, 22607 Hamburg, Germany}}
\affiliation{Dipartimento di Fisica, Universit\'{a} di Roma "La Sapienza'', P.le Aldo Moro 2, 00185 Roma, Italy}
\author{Beatrice Maroni}
\affiliation{Dipartimento di  Chimica, Sezione Chimica Fisica, INSTM (UdR Pavia), Universit\'a di Pavia, Viale Taramelli 16, 27100 Pavia, Italy}
\author{Carlo Marini}
\affiliation{European Synchrotron Radiation Facility, 6 Rue Jules Horowitz, BP220, 38043 Grenoble Cedex, France}
\author{Michael Hanfland}
\affiliation{European Synchrotron Radiation Facility, 6 Rue Jules Horowitz, BP220, 38043 Grenoble Cedex, France}
\author{Boby Joseph}
\affiliation{Dipartimento di Fisica, Universit\'{a} di Roma "La Sapienza'', P.le Aldo Moro 2, 00185 Roma, Italy}
\author{Paolo Postorino}
\affiliation{CNR-IOM and dipartimento di Fisica, Universit\'{a} di Roma "La Sapienza'', P.le Aldo Moro 2, 00185 Roma, Italy}
\author{ Lorenzo Malavasi}
\affiliation{Dipartimento di Chimica, Sezione Chimica Fisica, INSTM (UdR Pavia), Universit\'a di Pavia, Viale Taramelli 16, 27100 Pavia, Italy}
\date{\today}

\begin{abstract}
We present structural studies of V$_{1-x}$Cr$_x$O$_2$ (pure, 0.7\% and 2.5\% Cr doped) compounds at room temperature in a diamond anvil cell for pressures up to 20 GPa using synchrotron x-ray powder diffraction. 
All the samples studied   show a persistence of the monoclinic $M_1$ symmetry between 4 and 12 GPa. Above 12 GPa,  the monoclinic $M_1$ symmetry changes to isostructural $M_x$ phase (space group $P2_1/c$) with a significant anisotropy in lattice compression of the $b$-$c$ plane of the $M_{1}$ phase. 
This behavior can be reconciled invoking the pressure induced charge-delocalization.

Journal reference : {\it Phys. Rev. B} \href{http://link.aps.org/doi/10.1103/PhysRevB.85.184108}{ 85, 184108 (2012)}
\end{abstract}

\maketitle

\section{Introduction}
The metal insulator transition (MIT) observed in most of
the vanadium oxides compounds belonging to the so-called Magn\'eli phase has attracted a lot of interest
\cite{perucchi09,morin59,longo70,goode71,marezio72,mott75,rice94,haverkort05,basov07,basov06,
arcang07,yao2010,eyert2011, belozerov2012,pashkin2011,qazilbash2011}.
Among these compounds VO$_2$ is probably the most investigated \cite{perucchi09}. On decreasing the temperature below 340 K, it
shows an abrupt MIT, characterized by a huge jump in the conductivity and a simultaneous structural transition
from rutile ($R$) metallic phase to monoclinic ($M_1$) insulating phase \cite{morin59}. In the $R$ phase, the
V-atoms, each surrounded by an oxygen octahedron, are equally spaced in linear chains along the $c$-axis and
form a body-centered tetragonal lattice \cite{marezio72}. The transition from the high-symmetry $R$ phase to the
low symmetry $M_1$ phase results mainly in a rearrangement in the V-chains. Structural data show indeed
the dimerization of the V-atoms in the chain and the tilting of the V-V pairs with respect to the $c$ axis
\cite{longo70}. The MIT in VO$_2$ has been a subject of intense study to pin-point the key players driving the
transition \cite{perucchi09}. Several scenarios have been proposed, an electron correlation-driven Mott transition
\cite{cao2010,basov07,mott75,rice94} , a structure-driven Peierls transition \cite{goode71,wentz94} or a
cooperation of these two mechanisms \cite{haverkort05,koethe06}. The understanding of VO$_2$ physics is even
more complicated due to the presence of two additional insulating monoclinic phases of this material, which
appear on applying uniaxial stress \cite{pouget75} or on doping with small amounts of acceptor (W or Nb)
\cite{mott75,reyes76} or donor (Cr) \cite{marezio72} impurities. In particular, on doping VO$_2$ with small amounts of Cr, $M_2$ and M$_3$ monoclinic phases (space group $C 2/m$) appear. These are characterized by
different arrangement of the V chains \cite{marezio72,pouget76}. In the $M_2$ structure the V pairing is
partially removed, \textit{ i.e.} one half of the V atoms dimerizes, and the other one forms zig-zag chains of
equally spaced atoms. The $M_3$ phase is an intermediate one in which the V-V pairs of the $M_2$ phase tilt and
the zig-zag chains start to pair.

We have recently investigated the effects induced by the application of high pressure (HP) on VO$_2$ based
systems using Raman and infrared spectroscopy \cite{perucchi09, arcang07, marini08,marini10}. Differently from the ambient
pressure case, where the metallic phase is found only in conjunction with the R structure, evidences of a new
room temperature metallic monoclinic phase have been found above 10 GPa \cite{arcang07,marini08,marini10}.
These findings opened to new experimental/theoretical queries about the HP structural arrangement of the V atoms
and its role in driving the MIT in the VO$_2$-systems. A HP structural study of pure and Cr-doped VO$_2$ is of
interest for a deeper understanding of the high pressure delocalization dynamics and for a description of the
Peierls-type distortion evolution. Here we report on a HP (0-20 GPa) synchrotron x-ray diffraction (XRD) study
carried out on V$_{1-x}$Cr$_x$O$_2$, where $x$=0.0\%, $0.7$\% and 2.5 \% under ambient pressure and temperature are respectively in monoclinic $M_1$, $M_3$, and $M_2$ phases. Regardless of the Cr content, all the investigated systems are found to be in monoclinic M$_1$ phase between 4 and 12 GPa. Above 12 GPa the
monoclinic $M_1$ phase transforms into an isostructural monoclinic $M_x$ phase (space group $P2_1/c$). In addition, our results reveal a significant anisotropy in lattice compression in the high pressure regime.
This behavior can be reconciled invoking the pressure induced charge-delocalization.

\begin{figure*}[t]
\includegraphics[width=17.5cm]{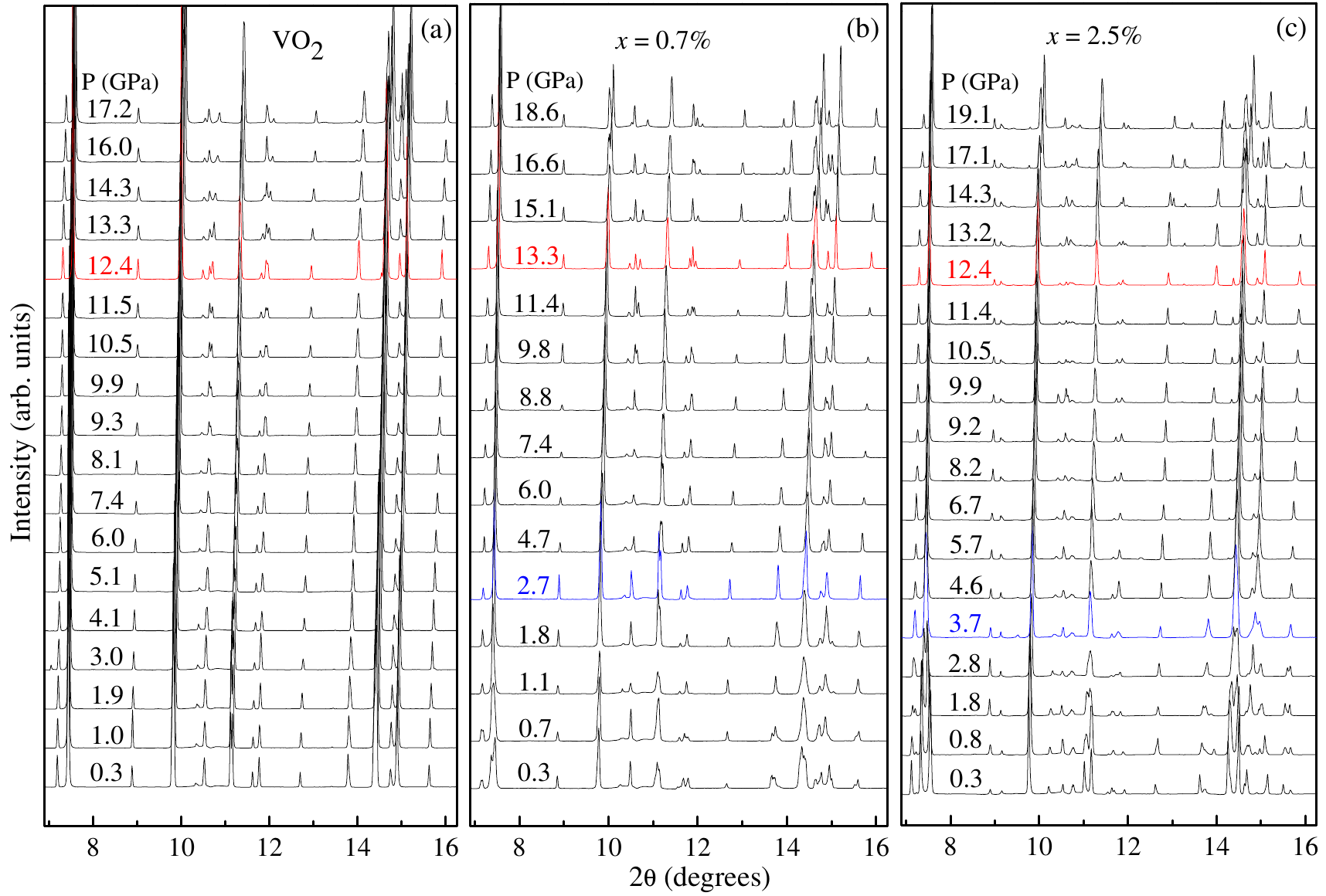}
\caption{(Color online) X-ray powder diffraction pattern series of V$_{1-x}$Cr$_x$O$_2$ (pure, 0.7\% Cr doped and 2.5\% Cr doped, panel (a), (b) and (c) respectively) compounds for several pressures. Red patterns mark the $M_1\rightarrow$M$_x$ transition pressure, the blue ones in (b) and (c) refer to respectively the low pressure $M_3\rightarrow M_1$ and $M_2\rightarrow M_1$ transitions.}
\label{fig1bis}
\end{figure*}

\section{Experimental}
VO$_2$ samples were prepared starting from proper amounts of V$_2$O$_3$ and V$_2$O$_5$ (Aldrich, purity $>$99.9\%) pressed to form a pellet and reacted at 1050 \textdegree C in argon gas flow for 12 hours. The Cr doped samples, (V$_{1-x}$Cr$_x$O$_2$ with $x$=0.025 and 0.007 respectively in the $M_2$ and $M_3$ monoclinic insulating phase at ambient conditions \cite{marezio72}), were prepared starting from equimolar amounts of V$_2$O$_3$ and V$_2$O$_5$ and adding proper stoichiometric amounts of Cr$_2$O$_3$. Notice that throughout the manuscript, we have used 'precentage format' ({\it i.e.} $x$=0.007 as 0.7\% and $x$=0.025 as 2.5\%) to denote the Cr concentration. In the Cr doped case, the mixture has then been fired at 1000 \textdegree C in argon atmosphere. From the above procedure, one can obtain small crystals of a few tens of microns. Preliminary phase purity checks were carried out on several of these obtained micron sized single-crystals through x-ray diffraction (XRD). Once the phase purity is assured, the single crystal pieces were crushed in a mortar to obtain the required powdered samples. These powdered samples were analysed using differential scanning calorimetery (DSC) and powder XRD. The refinement of the room temperature powder XRD patterns yielded $M_1$, $M_2$, and $M_3$ monoclinic structures for the $x$=0.0\%, 2.5\% and 0.7\% respectively with corresponding unit-cell parameters in excellent agreement with the reported values \cite{marezio72}. The DSC measurements revealed clear endothermic peaks for the $x$=0.0\%, 2.5\% and 0.7\% samples associated with the insulator-metal transition. The observed transition temperatures were in perfect agreement with the literature values \cite{morin59,marezio72}.

\begin{figure}[h]
\includegraphics[width=8.4cm]{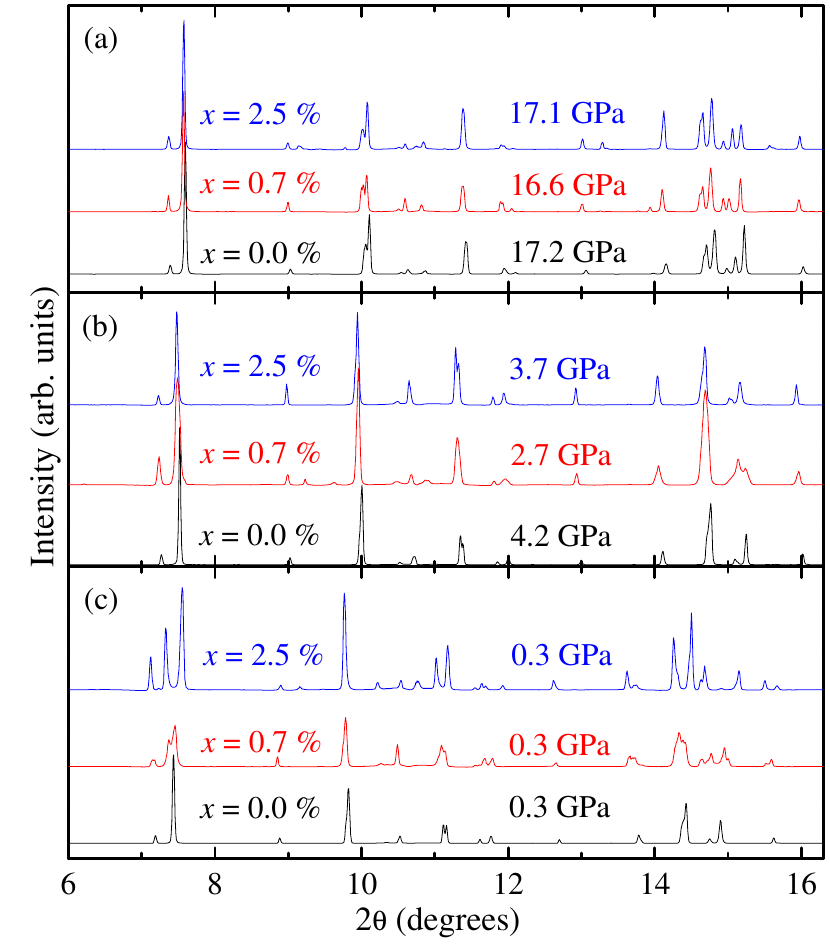}
\caption{(Color online) X-ray diffraction pattern of V$_{1-x}$Cr$_x$O$_2$ (pure, 0.7\% and 2.5\% Cr doped) compounds at several selected pressures. To facilitate a better comparison, the intensities of  the VO$_2$ patterns are reduced to one-third.}
\label{fig1}
\end{figure}

For HP measurements, the samples have been loaded in a membrane diamond anvil cell with a large angular aperture. Measurements were carried out in the pressure range 0 - 20 GPa. 
Helium was the selected pressure transmitting medium since it minimizes non-hydrostaticity effects for pressures even higher than those realized in the present work \cite{kenji01} and thus allows good quasi-hydrostatic conditions. The pressure was measured by means of ruby
fluorescence \cite{mao78}.
The monochromatic x-ray signal ($\lambda$=0.415 \AA) diffracted by the samples has been collected on a
MAR555 flat panel detector, located at a distance of $\sim$ 400 mm from the sample, in the ID09A beamline of
the European Synchrotron Radiation Facility, Grenoble (France). The diffraction parameters were carefully
determined using a silicon reference sample. XRD data have been analyzed using the FULLPROF \cite{fullprof}
Rietveld refinement package. Each pattern has been studied by modeling the structure according to the Rietveld
method and thus refining the structural information.

\section{Results and discussion}
\subsection{High pressure x-ray diffraction data}
\begin{figure}[h]
\includegraphics[width=8.4cm]{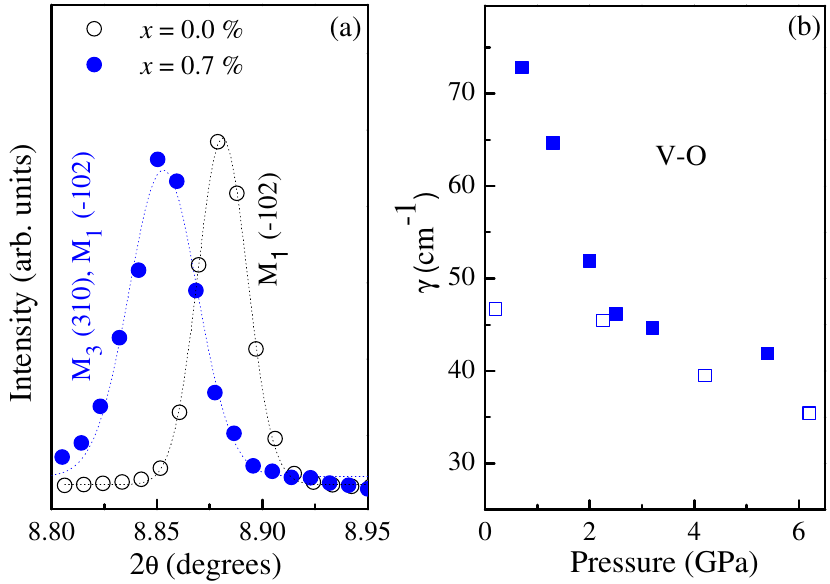}
\caption{(Color online) Comparison of a single Bragg reflection with comparable intensity (panel (a)) and Raman phonon linewidths (panel (b)) of the modes at $\sim$640 cm$^{-1}$ (squares) pressure dependence for pure (empty symbols) and 0.7 \% Cr-doped VO$_2$ (filled symbols) \cite{marini08, marini10}.}
\label{Raman}
\end{figure}

XRD   patterns   for   the V$_{1-x}$Cr$_x$O$_2$ (pure, 0.7\% and 2.5\% Cr doped) samples at several pressures are shown in Fig. 1.   
For facilitating a better comparison of the XRD data at different 
pressure regimes (low, medium, and high), in Fig.  \ref{fig1}, we present the XRD pattern at three selected pressures for the three compounds studied. 
Data collected at the lowest pressures [Fig. \ref{fig1} (c)] show clearly different peak patterns in accordance with the
different symmetries expected for V$_{1-x}$Cr$_x$O$_2$ at ambient pressure \cite{marezio72}. Data refinements of
the diffraction patterns collected at $P$ = 0.3 GPa show that the crystal structures of the pure, 0.7\% and 2.5\% Cr doped samples belong to respectively the $M_1$, $M_3$, and $M_2$ monoclinic phases as expected. As can be inferred by
comparing the diffraction patterns reported in the different panels of Fig. \ref{fig1} (see also Fig. 1), the three samples show
different behaviors over the low-pressure range (0-4 GPa approximately) but a convergence towards a common
structure and a common trend is observed in the high-pressure regime (P$>$4 GPa). In particular, pure VO$_2$
sample shows a rather regular pressure dependence up to about P$^*$=12 GPa.  From Fig. \ref{fig1bis}(a), it can be clearly observed how the Bragg peaks shift towards high angles owing to lattice compression without any modification of the overall peak pattern up to P$^*$. Pressure induced structural transitions in the low pressure regime (P$<$ 4 GPa) are instead observed for the two Cr-doped samples [Figs. \ref{fig1bis} (b) and (c)]. Data refinement of the diffraction patterns clearly shows the transition of the two doped samples towards the $M_1$ phase, characteristic of the pure compound, on increasing the pressure: the $M_3 \rightarrow M_1$ transition of the 0.7\% Cr doped sample occurs at 2.7 GPa whereas, the $M_2 \rightarrow M_1$ transition of the 2.5\% doped at 3.7 GPa [blue in Figs. \ref{fig1bis} (b) and (c)]. 
The evolution of the crystal-structures can be followed looking at the three Bragg peaks at around 7.3\textdegree, characteristic of the $M_2$ and $M_3$ phases which transform into the doublet of the $M_1$ phase (see Figs. \ref{fig1bis} and \ref{fig1}). Similarly, the sequence of peaks located at 11.0\textdegree$ $ and 14.5\textdegree$ $ which merge into single peaks for P$>$ 4 GPa. The three samples, regardless the extent of Cr doping, show the same diffraction peak pattern and the same pressure behavior above the transiton pressure, 2.7 GPa for $x$=0.7\% and  3.7 GPa for $x$=2.5\% Cr doped, up to P$_{max}$ = 20 GPa (panels (b) and (c) of Fig. 1; see also Fig. \ref{fig1}). This finding is consistent with the observed pressure dependence of the Raman-active phonons reported in Ref. [\onlinecite{marini08}] where the same three samples here investigated exhibit peculiar dependence over the low-pressure range and actually converge into a common one in the high-pressure regime.

\begin{figure}[h]
\includegraphics[width=8.4cm]{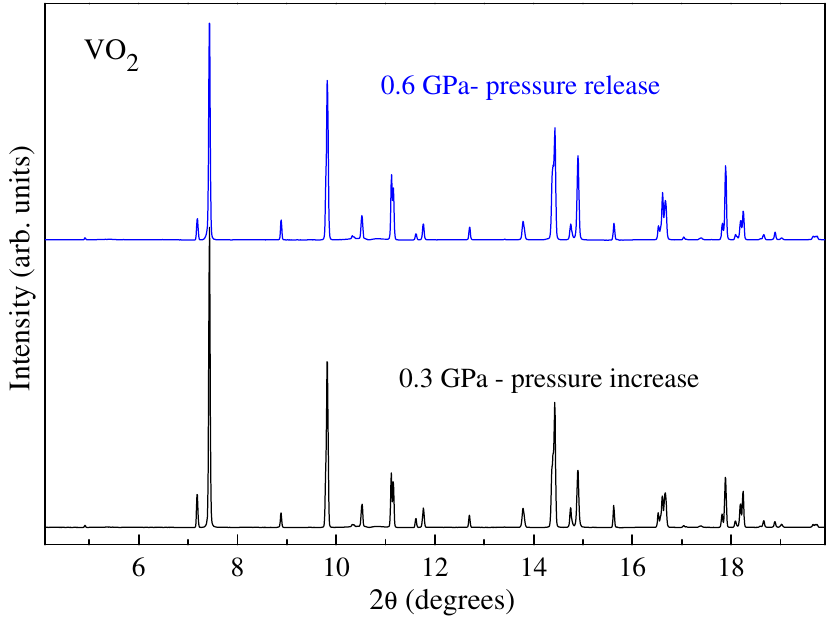}
\caption{(Color online) X-ray powder diffraction patterns for VO$_2$ below 1 GPa before and after the pressure cycle. The two patterns are identical revaealing the reversibility of the pressure induced phase transitions.}
\label{fig1ter}
\end{figure}

At pressures above P$^*$ (red patterns in Fig. \ref{fig1bis}), subtle modifications occur in the diffraction patterns of all the three samples such as
the splitting of the peak at 10\textdegree~ and the rearrangements of the peaks around 15\textdegree.
Nevertheless no symmetry changes are needed to model the diffraction data and a monoclinic phase, we named as $M_x$ according to Ref. [\onlinecite{marini08}], isostructural to $M_1$ (space group $P2_1/c$), describes the
structure of all the samples above P$^*$.

\begin{figure}[h]
\includegraphics[width=8.4cm]{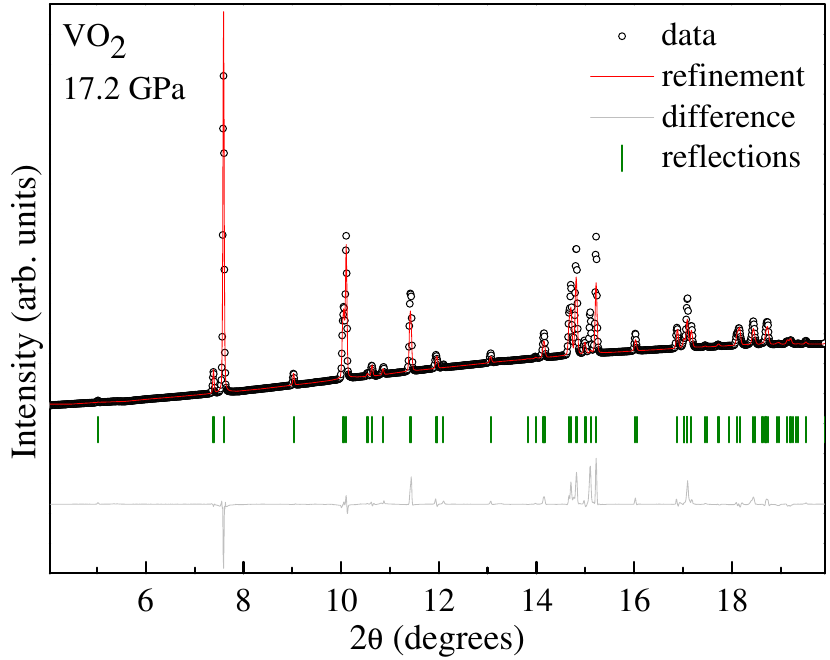}
\caption{(Color online) Results of the Rietveld refinement for VO$_2$ in the M$_x$ phase at 17.2 GPa.}
\label{fig3bis}
\end{figure}

Moreover, as to the $x=0.7\%$ sample, it can be noticed that Bragg peaks shown in Fig. \ref{fig1} at P = 0.3 GPa are apparently broader than those of other samples. This is quite clear looking at Fig. \ref{Raman} (a) where the peak around 8.90\textdegree~ is shown for the pure and $0.7\%$ doped samples at P=0.3 GPa. This broadening is due to the phase coexistence at low pressure-regime between the $M_3$ and $M_1$ phases in the 0.7\% sample. In fact XRD refinements show the Bragg peak around 8.90\textdegree~ has contribution from both (310) of $M_3$ and (-102) of $M_1$ phases. The effect of this co-existence is also evident in the Raman results. 
The pressure dependence of the phonon line width, $\gamma$, of the V-O Raman-active modes is shown Fig. \ref{Raman} (b) for the pure and the 0.7\% doped sample over 0-6 GPa range. Within a phase separation scenario
the observed progressive reduction of the $\gamma$ for the Cr-doped sample can be ascribed to a pressure induced
reduction of the $M_3$ phase in favor of the $M_1$.
This is also an indication of the first order nature of the $M_3$ to $M_1$ transition.
We also collected few diffraction patterns on releasing the pressure to verify the reversibility of the observed structural phase transitions. As can be observed from Fig. \ref{fig1ter}, where the powder XRD pattern for pure VO$_2$ are compared before and after the pressure ramp, the initial structure is found to be recovered upon the pressure release.  Thus the results of these measurements show that the pressure induced structural phase transitions are reversible.

\subsection{Pressure dependence of lattice constants}
\begin{figure}[h]
\includegraphics[width=8.4cm]{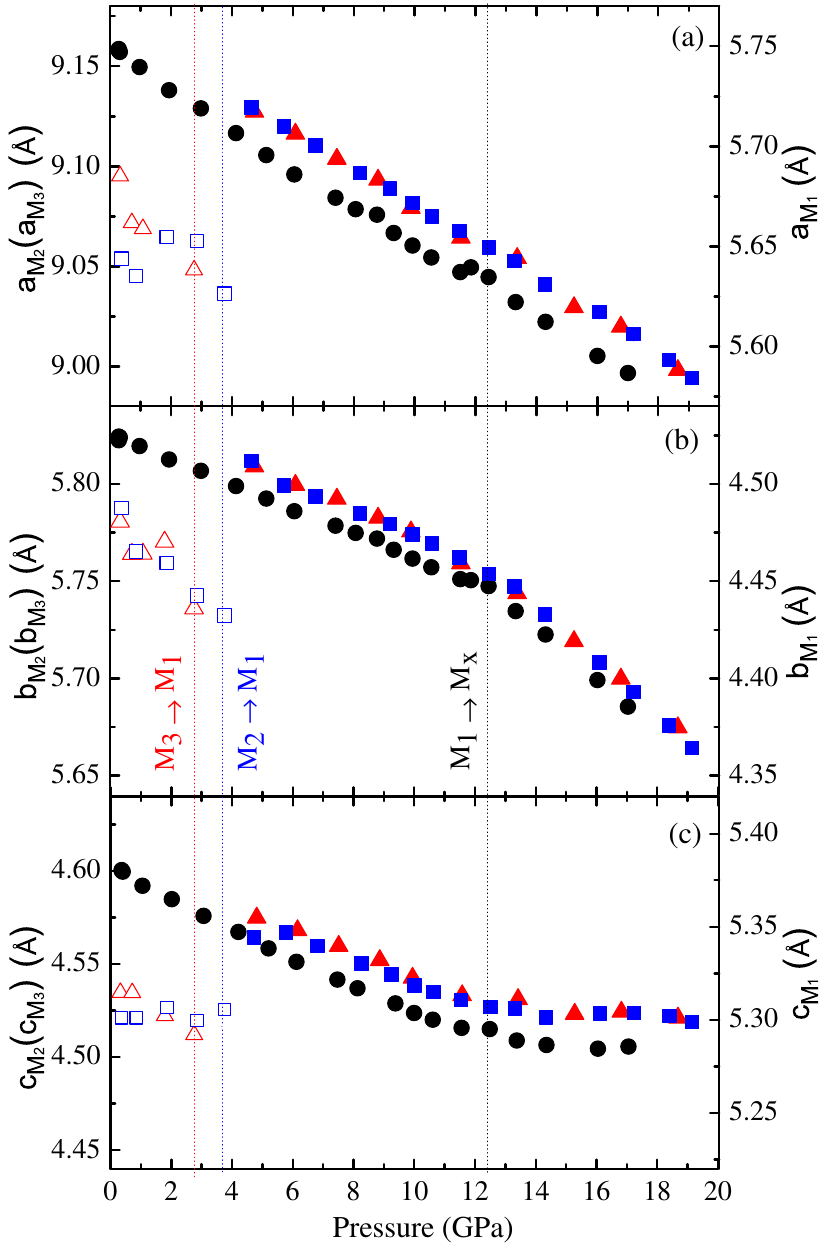}
\caption{(Color online) Pressure dependence of $a$, $b$, and $c$ lattice parameters [panels (a), (b) and (c)] for the investigated
samples (circles, triangles, and squares for pure, 0.7\% and 2.5\% Cr doped compounds respectively). The empty symbols refer to
$M_2$ and $M_3$ cells (left scale), filled symbols refer to the $M_1$ and $M_x$ phases (right scale). Vertical
dotted lines mark the transition pressures.} \label{fig3}
\end{figure}

Refinements of the XRD patterns at different pressures permits to obtain the pressure dependence of the lattice parameters, which is shown in  Fig. \ref{fig3}. An example of data refinement ( the VO$_2$ pattern at 17.2 GPa), is shown in Fig. \ref{fig3bis}. The quality of data allows indeed for a good estimate of the lattice parameters, whereas it is not sufficient to follow the precise evolution of the atomic positions in the high pressure regime.  For the sake of completeness, the obtained parameters (e.g., lattice constants and atomic positions) from the XRD Rietveld refinement analysis are given as tables in the supplementary information file.

It is worth to notice that when changing from the $M_2$ or the $M_3$ conventional cell to the $M_1$ cell, the geometrical meaning of cell axes changes according to a non-trivial relation (see Ref.  \onlinecite{marezio72}). 
At low pressure (P$<$ 4 GPa) the Cr-doped samples exhibit a rather scattered behavior (mainly the $x$ = 0.7\% sample) of all
the lattice constants, which highlights an embedded instability of low symmetry structures M$_2$ and M$_3$ that
evolve towards $M_1$. $M_1$ ($P2_1/c$) and $M_2-M_3$ ($C 2/m$) structures show different arrangements of the unit cells, with respectively 12 and 24 atoms (doubling of the $M_1$ unit cell with respect to $M_2$($M_3$) \cite{marezio72,pouget75,pouget76}) per unit-cell. As mentioned above, on changing the space group the geometrical meaning of the  lattice parameters is modified \cite{pouget75} and  as consequence a deceptive discontinuity in the pressure dependences shown in Fig. \ref{fig3} is observed. The discontinuity indeed disappears once dealing with the normalized unit cell volume (see Section III C: Equation of state). 
Let us now focus on the high pressure regime, P $>P^*$. In this case we basically do not consider differences among
the three samples anymore. Looking at Fig. \ref{fig3} it is evident that although the $a$ axis exhibits a regular continuous decrease with pressure, a strong anisotropic compression of the $b$-$c$ plane (of the $M_1$ phase) occurs. As a matter of fact, a substantial reduction of the {\it axial bulk modulus} of the $b$ lattice parameter, defined as $\Delta P/\Delta b$, from $\sim$172 GPa/\AA$ $ to $\sim$95 GPa/\AA, with a simultaneous hardening of the $c$ axis is observed above P$^*$. This remarkable anisotropic compression
within the same crystal structure suggests a rearrangement of charge density, which can be related to
the pressure-induced insulator-to-metal transition observed in infrared (IR) spectra reported in Refs. [\onlinecite{arcang07,marini08}] above 10 GPa.
From a qualitative point of view a modification in the electron density, $n$, should imply a different lattice response under volume compression. As a matter of fact,  the bulk modulus is found to be proportional to $n^{5/3}$ in a free electron gas \cite{mahan00}. Since above P$^*$ the {\it axial bulk modulus} increases along $c$ and decreases along $b$ a charge-transfer from the $b$ axis towards the $c$ on entering the metallic phase can be inferred.

\subsection{Equation of state}

\begin{figure}[h]
\includegraphics[width=8.4cm]{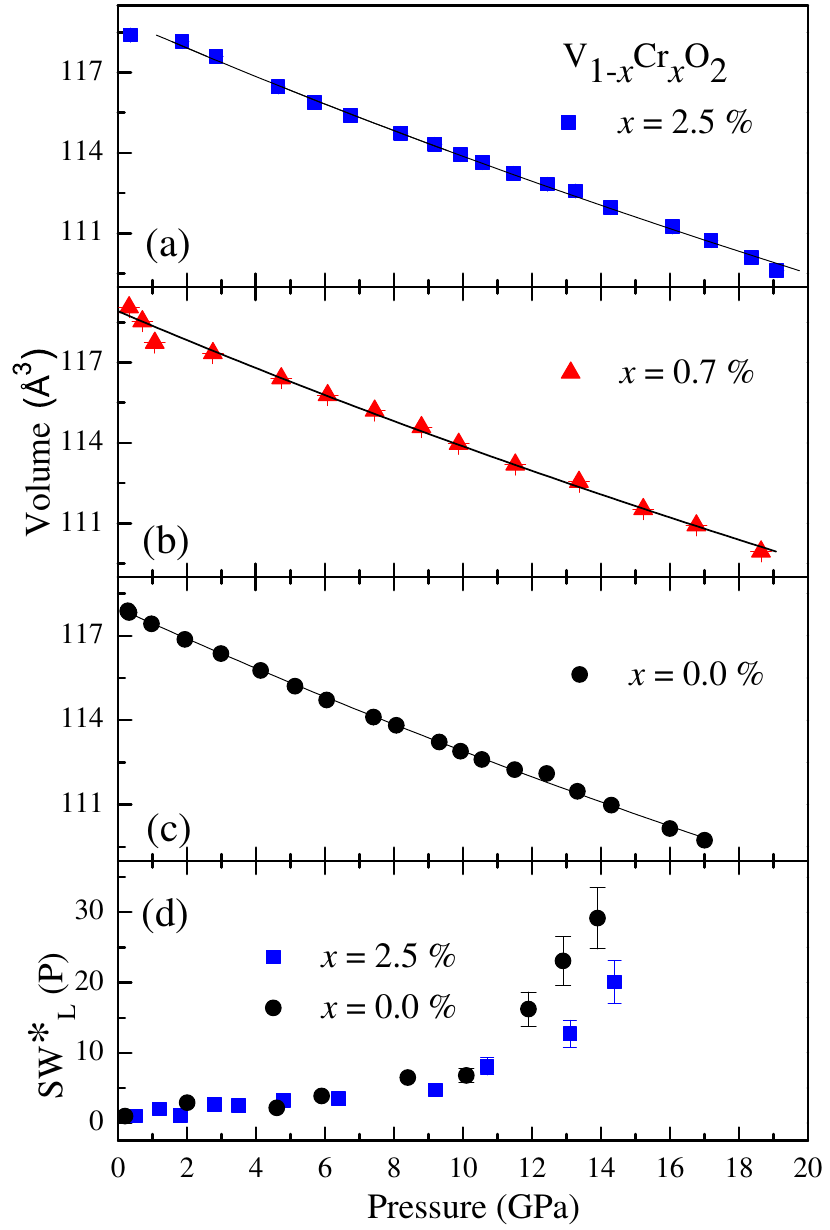}
\caption{(Color online) Pressure dependence of the unit-cell volume for 2.5\% Cr doped (a), 0.7\% Cr doped (b) and pure (c) samples (symbols) together with the Birch-Murnaghan fits. Panel (d) shows the infrared spectral weight transfer as a function of pressure taken from Refs. [\onlinecite{arcang07,marini08}] for pure and 2.5\% doped samples.}
\label{fig4}
\end{figure}

The unit-cell volume derived from the pressure dependence of the lattice parameters reported above is shown in Fig. \ref{fig4}.
\begin{table}[h]
\begin{tabular}{c | c c}
\hline\hline
V$_{1-x}$Cr$_x$O$_2$ & $V_0$ (\AA$^3$) & $K_0$ (GPa)\\
\hline
$x$ = 0.0 \% & $117.985\pm0.042$ & $141.7\pm1.3$\\
$x$ = 0.7 \% & $118.92\pm0.11$ & $144.0\pm3.1$\\
$x$ = 2.5 \% & $119.021\pm0.086$ & $141.1\pm2.2$\\
\hline\hline
\end{tabular}
\caption{Bulk modulus,  $K_0$ and conventional cell volume, $V_0$  obtained from the Birch-Murnaghan model fit to the pressure vs unit-cell volume data for  V$_{1-x}$Cr$_x$O$_2$ shown in Fig. \ref{fig4} panels (a)-(c).}
\label{table1}
\end{table}
As mentioned above, we notice that the deceptive discontinuities of the lattice parameters on entering the M$_1$ phase  (see Fig. \ref{fig3}) disappear once dealing with unit cell volumes. Moreover, no significant discontinuities are observed on entering the metallic phase ruling out the onset of simultaneous structural transitions. The pressure dependence of the infrared spectral weight taken from Ref.  [\onlinecite{arcang07} and \onlinecite{marini08}] is also shown for sake of comparison in Fig.  \ref{fig4} (panel (d)). The pressure dependence of the unit-cell volume shown in  panels (a), (b) and (c) of Fig.  \ref{fig4} can be well described by the Birch-Murnaghan (BM) equation of state \cite{birch47}:
\begin{eqnarray}
\lefteqn{P(V)=\frac{3}{2} K_0\bigg[\bigg(\frac{V_0}{V}\bigg)^{7/3}-\bigg(\frac{V_0}{V}\bigg)^{5/3}\bigg]\times}\nonumber\\
&& {} \times\bigg\{1+\frac{3}{4}(K_0'-4)\bigg[\bigg(\frac{V_0}{V}\bigg)^{2/3}-1\bigg]\bigg\}
\end{eqnarray}
using the $P$=0 unit-cell volume $V_0$, and the bulk modulus $K_0$ as free fitting parameters and setting the pressure derivative of the bulk modulus $K_0'$ to four \cite{lei07,rivad09}. Best fit parameters are reported in Table \ref{table1}  and best fit curves are shown as solid lines in panels (a), (b) and (c) of Fig. \ref{fig4}.
The present results demonstrate that the high pressure metallic phase is apparently a structural phase common to all the compounds studied regardless the different ambient pressure structural symmetries. The fact that the metallic $M_x$ phase has the same symmetry of the insulating $M_1$ phase implies that the former preserves the V dimers up to the maximum pressure. The only relevant difference of the metallic $M_x$ phase from the insulating $M_1$ is the remarkably different response of the lattice upon volume compression (Figs. \ref{fig3} and \ref{fig4}). In addition, bearing in mind that the extent of Peierls distortion is increasing on going from the $M_2$ to the $M_3$ to the $M_1$ phase indicate that the lattice compression in these systems, surprisingly appears to enforce the Peierls distortion. This peculiar behavior can represent a benchmark for any future theoretical work aimed at addressing the physics of these highly correlated systems.

\section{Summary and conclusions}
In conclusion, we have investigated structural evolution as a function of pressure of V$_{1-x}$Cr$_x$O$_2$ (pure, 0.7\% and 2.5\% Cr doped) compounds. As pressure is raised, the 0.7\% and 2.5\% doped compounds showed a structural
transition to the $M_1$ phase respectively at P$\sim$ 2.7 and 3.7 GPa. On further increasing the pressure, the
three systems remain in a $M_1$ symmetry, with a monotonic decrease of the unit-cell parameters until a
threshold pressure, P$^*=12$ GPa. Above P$^*$, all the samples show a remarkable anisotropic compression without
any change in crystal symmetry, which remains as M$_1$. In this high pressure regime (P$>$P$^*$), the $b$ axis
becomes more compressible while the $c$ axis hardens indicating a non-trivial rearrangement of charge densities
occurring over the $b$-$c$ plane. The onset of the anisotropic compression occurs at a pressure quite close to
the metallization boundary identified in the high pressure infra-red spectra \cite{arcang07}. This behavior can
be thus regarded as a structural signature of the metallization transition occurring in VO$_2$ and in the
Cr-doped compounds.

At ambient pressure, the activation of the Peierls distortion along the V-V chains is regarded as the major microscopic mechanism for the onset of the low temperature insulating phase, although experimental evidences suggest also the relevance of strong electronic correlations.  On applying the pressure, our present findings together with the optical data reported in Refs. [\onlinecite{arcang07, marini08, marini10}], reveal the simultaneous coexistence of a Peierls distorted monoclinic phase together with an enhancement of the metallicity of the system. The electronic modifications in the high pressure regime (P$>$P$^*$) occur without a  clear, simultaneous symmetrization of  the V-V chains and regardless the different extent of the Cr-induced distortion.  Present results thus demonstrate the possibility of a MIT mainly driven by electron-electron correlation and these results represent a severe bench mark for any theoretical modeling aimed at describing the electronic and structural modifications of VO$_2$ induced by temperature and lattice compression.

\section*{Acknowledgments}
The authors thank ESRF, Grenoble, for the beamtime allocation for the high pressure XRD measurements.

\newpage

\section*{Supplementary information}
\subsection*{Refinement  results  for the $x$ = 0.0\%, sample  at  three selected pressures.}

\begin{figure}
\includegraphics[width=14.5cm]{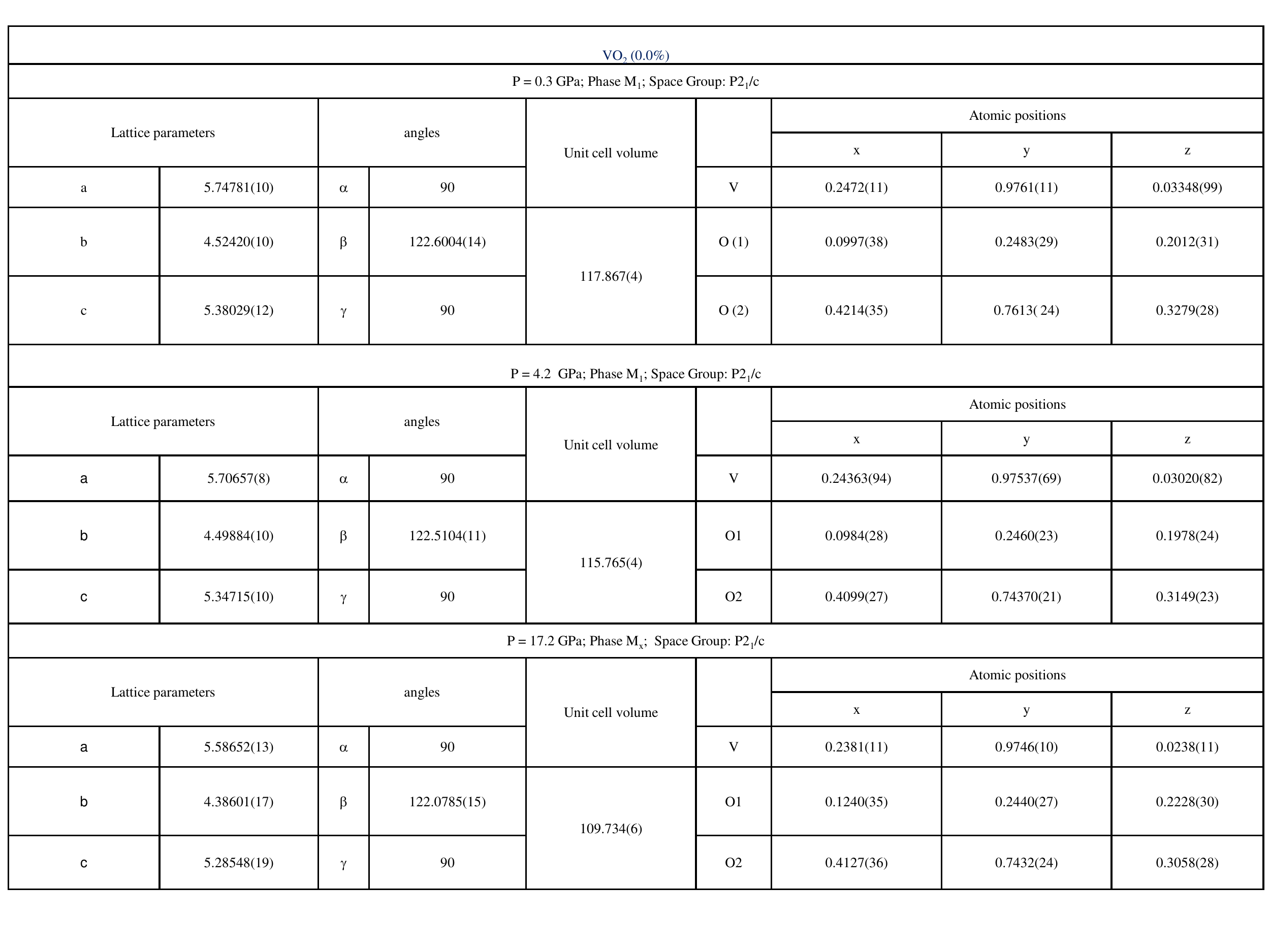}
\label{fig-SI-1}
\end{figure}

\newpage

\section*{Supplementary information}
\subsection*{Refinement  results  for the   0.7\% sample  at  three selected pressures.}

\begin{figure}
\includegraphics[width=14.5cm]{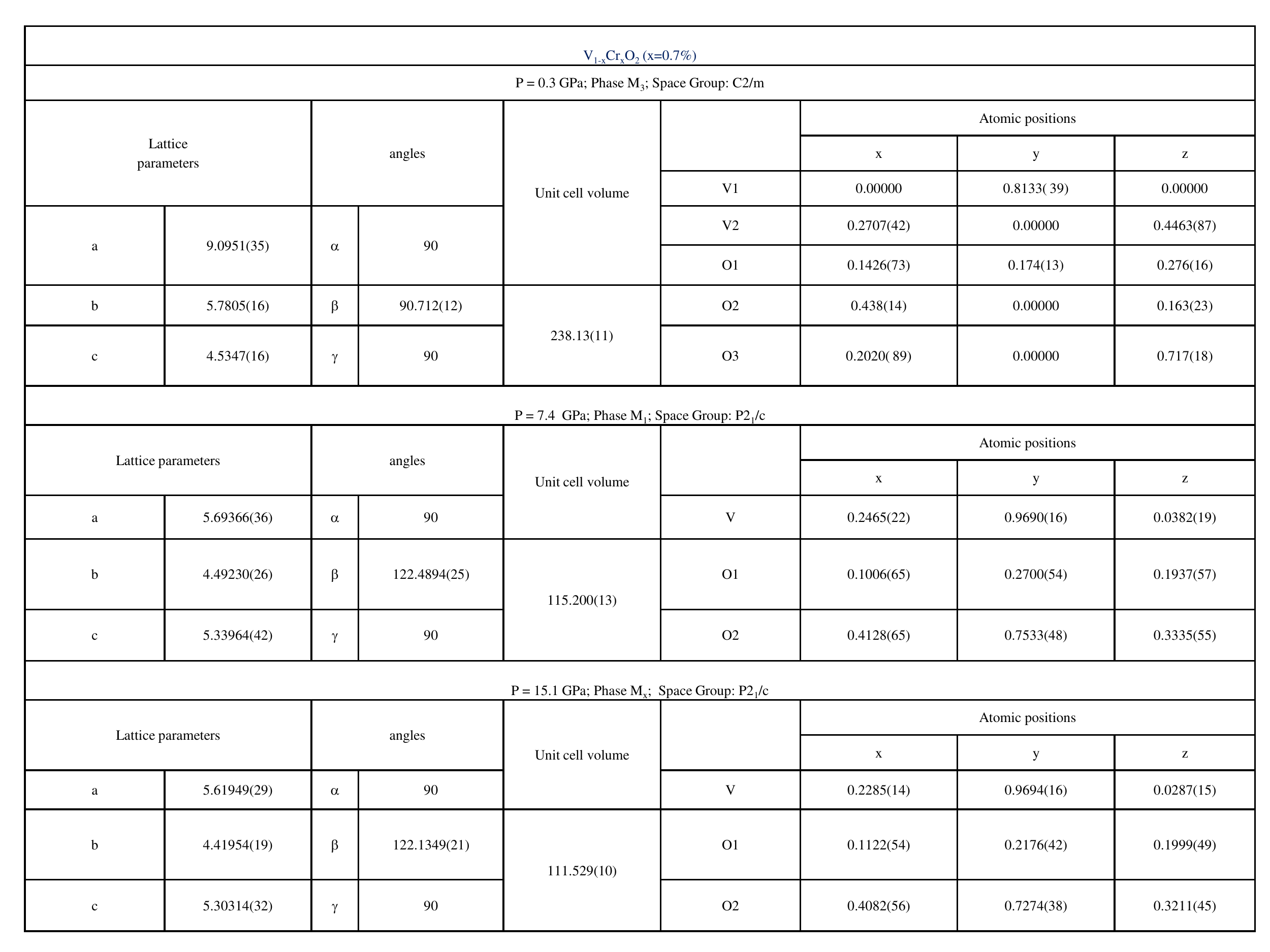}
\label{fig-SI-2}
\end{figure}

\newpage

\section*{Supplementary information}
\subsection*{Refinement  results  for the2.5\% sample  at  three selected pressures.}
\begin{figure}
\includegraphics[width=14.5cm]{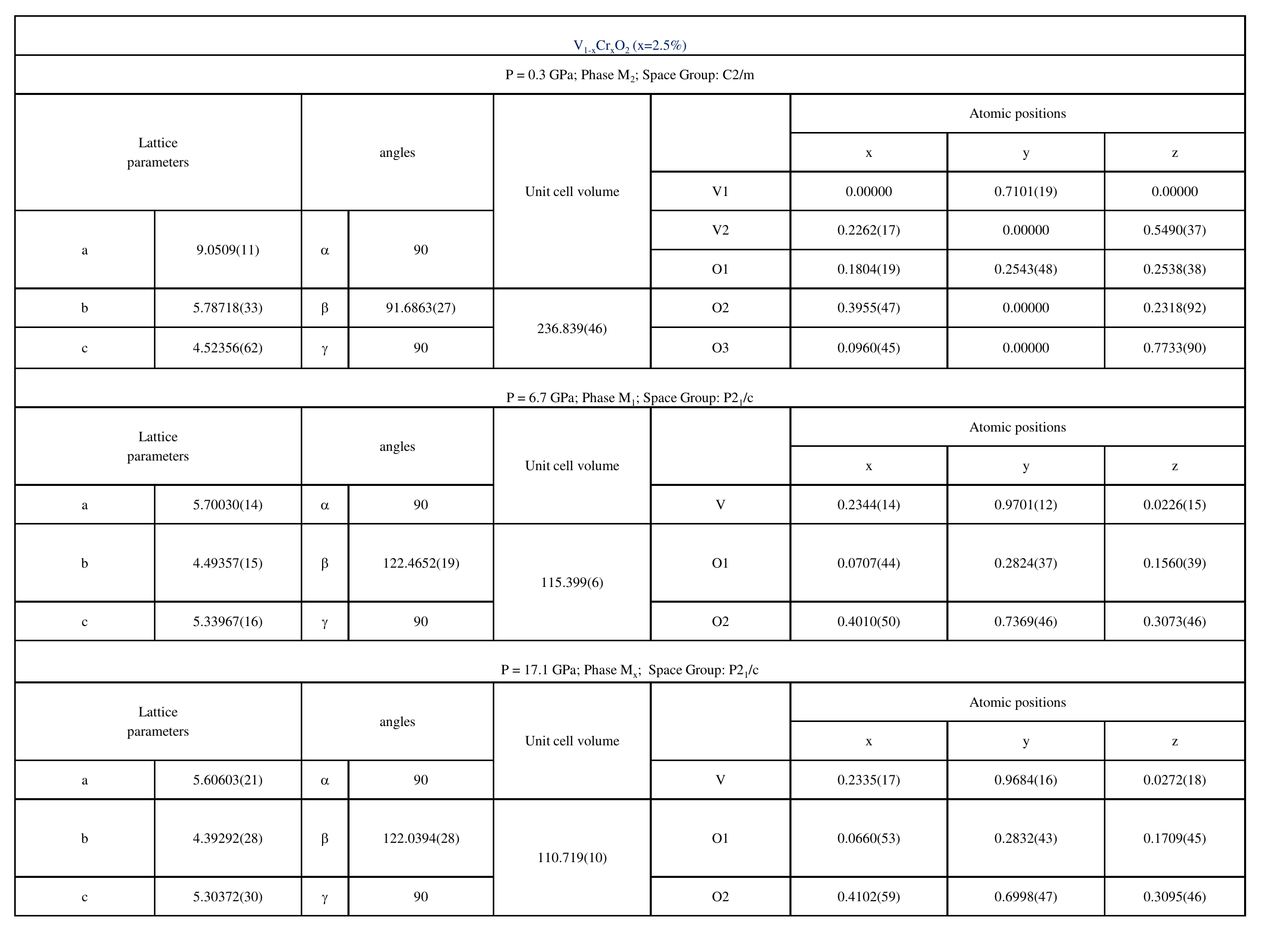}
\label{fig-SI-3}
\end{figure}

\newpage
\section*{Supplementary information}
\subsection*{Pressure behavior of the monoclinic $\beta$ angle for pure (circles), 0.7\% Cr doped (triangles), and 2.5\% doped (squares) VO$_2$ in the $M_1$ and $M_x$ phases.}

\begin{figure}
\includegraphics[width=8.5cm]{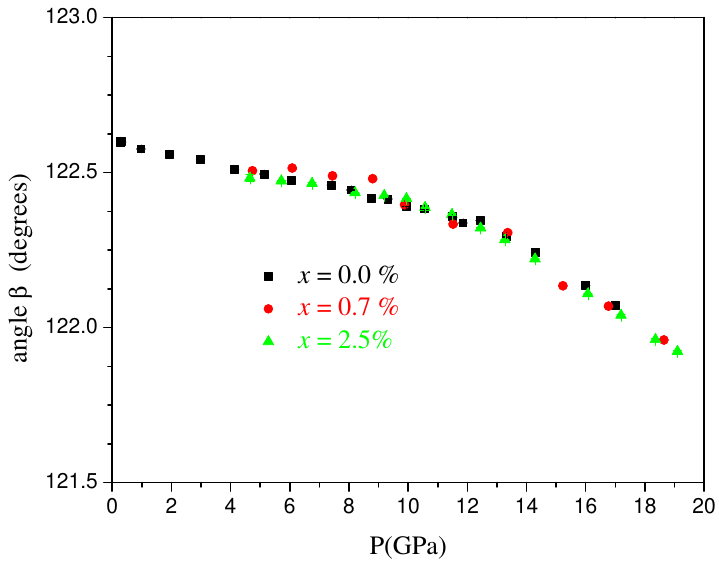}
\label{fig-SI-3}
\end{figure}

\end{document}